\documentclass[%
    aps,
    prd,
    reprint,
    10pt,
    nofootinbib,
    superscriptaddress,
    longbibliography,
    amsfonts,
    amssymb,
    amsmath,
    preprintnumbers,
    noshowpacs,
    noshowkeyws,
    tightenlines,
    floats,
    notitlepage,
    final,
    letterpaper,
    oneside,
    twocolumn,
    ]{revtex4-2}
\pdfoutput=1
\usepackage[utf8]{inputenc}
\usepackage[T1]{fontenc}
\usepackage[british]{babel}
\usepackage[british,cleanlook,printdayon]{isodate}
\usepackage[Symbolsmallscale]{upgreek}
\usepackage{mathtools}
\usepackage[protrusion,tracking,kerning,spacing,final,babel]{microtype}
\usepackage{physics}
\usepackage[dvipsnames]{xcolor}
\usepackage{graphicx}
\usepackage[final]{hyperref}
\usepackage[normalem]{ulem}
\usepackage{subfigure}
\hypersetup{%
    colorlinks=true,
    citecolor=MidnightBlue,
    linkcolor=MidnightBlue,
    urlcolor=MidnightBlue
    }%
\usepackage{cleveref}
\usepackage{nth}
\usepackage{orcidlink}

\begin{document}
%\title{Measuring gravitational-wave background anisotropies with\\ground-based interferometers in the presence of shot noise}
%\title{Unbiased inference of shot-noise-dominated gravitational-wave background anisotropies}
%\title{Unbiased inference of gravitational-wave background anisotropies\\with ground-based interferometers}
%\title{Unbiased mapping of the gravitational-wave sky from noisy data}
\title{Unbiased estimation of gravitational-wave anisotropies from noisy data}

\author{Nikolaos~Kouvatsos\,\orcidlink{0000-0002-5497-3401}}
\email{nikolaos.kouvatsos@kcl.ac.uk}
\affiliation{Theoretical Particle Physics and Cosmology Group, Physics Department, King's College London, University of London, Strand, London WC2R 2LS, UK}

\author{Alexander~C.~Jenkins\,\orcidlink{0000-0003-1785-5841}}
\email{alex.jenkins@ucl.ac.uk}
\affiliation{Department of Physics and Astronomy, University College London, London WC1E 6BT, UK}

\author{Arianna~I.~Renzini\,\orcidlink{0000-0002-4589-3987}}
\email{arenzini@caltech.edu}
\affiliation{LIGO Laboratory, California Institute of Technology, Pasadena, CA 91125, USA}
\affiliation{Dipartimento di Fisica “G. Occhialini”, Universit\`a  degli Studi di Milano-Bicocca, Piazza della Scienza 3, 20126 Milano, Italy}
\affiliation{INFN, Sezione di Milano-Bicocca, Piazza della Scienza 3, 20126 Milano, Italy}

\author{Joseph~D.~Romano\,\orcidlink{0000-0003-4915-3246}}
\email{joseph.romano@utrgv.edu}
\affiliation{Department of Physics and Astronomy, University of Texas Rio Grande Valley, One West University Boulevard, Brownsville, TX 78520, USA}

\author{Mairi~Sakellariadou\,\orcidlink{0000-0002-2715-1517}}
\email{mairi.sakellariadou@kcl.ac.uk}
\affiliation{Theoretical Particle Physics and Cosmology Group, Physics Department, King's College London, University of London, Strand, London WC2R 2LS, UK}

\date{\today}
\preprint{KCL-PH-TH/2023-67}

\begin{abstract}
    One of the most exciting targets of current and future gravitational-wave observations is the angular power spectrum of the astrophysical GW background.
    This cumulative signal encodes information about the large-scale structure of the Universe, as well as the formation and evolution of compact binaries throughout cosmic time.
    However, the finite rate of compact binary mergers gives rise to \emph{temporal shot noise}, which introduces a significant bias in measurements of the angular power spectrum if not explicitly accounted for.
    Previous work showed that this bias can be removed by cross-correlating GW sky maps constructed from different observing times.
    However, this work considered an idealised measurement scenario, ignoring detector specifics and in particular noise contributions.
    Here we extend this temporal cross-correlation method to account for these difficulties, allowing us to implement the first unbiased anisotropic search pipeline for LIGO-Virgo-KAGRA data.
    In doing so, we show that the existing pipeline is biased \emph{even in the absence of shot noise}, due to previously neglected sub-leading contributions to the noise covariance.
    We apply our pipeline to mock LIGO data, and find that our improved analysis will be crucial for stochastic searches from the current observing run (O4) onwards.
\end{abstract}

\maketitle

%%%%%%%%%%%%%%%%%%%%%%%%%%%%%%%%%%%%%%%%%%%%%%%%%%%%%%%%%%%%%%%%%%%%%%%%%%%%%%
\section{Introduction}
During its first three observing runs, the LIGO-Virgo-KAGRA Collaboration detected dozens of gravitational-wave (GW) signals from binary black hole coalescences~\cite{PhysRevX.9.031040, PhysRevX.11.021053, 2021arXiv211103606T}, with the fourth run currently ongoing. This number is expected to increase to several thousand by the end of the fifth observing run~\cite{KAGRA:2013rdx}. As a result, interest is now shifting to other, undetected observables, like the GW background (GWB)---a persistent signal made up of numerous superimposed GW sources thoughout the history of the Universe~\cite{1997rggr.conf..373A, Christensen_2019}. With the recent strong evidence for a GWB in the nanohertz band provided by pulsar timing arrays~\cite{EPTA_InPTA, NANOGrav, PPTA}, we can expect exciting developments in the field within the next few years.

While the GWB can originate from many different early-universe cosmological mechanisms~\cite{Caprini_2018}, the astrophysical GW background (AGWB)~\cite{Tania_Regimbau_2011} arising from the combined emission of compact binary coalescenses (CBCs) is expected to be the dominant component. In the past years, significant research has been devoted to probing the anisotropies in the AGWB, as these could provide us with valuable information with regards to the large-scale structure (LSS) of the Universe and, in particular, galaxy clustering~\cite{Contaldi:2016koz,Cusin:2017fwz,PhysRevLett.120.231101,PhysRevD.98.063501,PhysRevLett.122.111101,Cusin:2019jpv,Capurri:2021zli}.

Much like other cosmological observables such as the cosmic microwave background (CMB), the anisotropies in the AGWB are typically probed by means of their angular power spectrum, $C_\ell$. A significant challenge in the case of the AGWB is that the finite rate of CBCs leads to temporal shot noise, which can significantly bias inferences of the angular power spectrum of the AGWB~\cite{PhysRevD.100.063508,Mukherjee:2019oma,Muk2023}. It is essential that this bias be properly accounted for in order to calculate the true angular power spectrum.
Reference~\cite{Jenkins:2019nks} introduced a technique to remove this shot noise bias by performing a cross-correlation of GW sky maps produced from nonoverlapping epochs within the full dataset. However, this work did not consider limitations related to GW detector noise and interferometer orientation. In this paper, we show both in theory and in practice how the optimal $C_\ell$ estimator first introduced in Ref.~\cite{Jenkins:2019nks} does indeed compute the correct angular power spectrum even after one accounts for the aforementioned effects, unlike the estimator currently used by both the LVK Collaboration and other authors~\cite{Renzini2019A, Renzini2019B, PhysRevD.100.062001, directional_search} to probe the angular power spectrum of the AGWB~\cite{Thrane:2009fp}.

As a by-product of this analysis, we show that the existing search pipeline is biased \emph{even in the absence of shot noise}, due to sub-leading contributions to the covariance matrix.
These contributions are sourced by the GWB signal itself, which is much weaker than the noise power in the detectors, and is thus usually neglected.
However, as we show explicitly below, these terms can be significantly larger than the angular power spectrum one is searching for, and therefore bias the $C_\ell$ estimator by more than $100\%$.
The method we describe here is robust to these errors, and is thus the first unbiased search pipeline for GWB anisotropies with LVK data.

The paper is organised as follows. In Sec.~\ref{theory}, we show how the current $C_\ell$ estimator is biased with contributions from detector noise, shot noise and the signal itself, while the optimal estimator is by construction unbiased. Section~\ref{mock} describes the anisotropic analysis we perform on mock data to validate our theoretical predictions, for detector noise sampled from O3 and LIGO A+ sensitivity. We also make predictions for the currently ongoing O4 observing run. We conclude in Sec.~\ref{summary} by discussing the importance of our findings for current and increased detector sensitivity, and providing possible future extensions of our work.

%%%%%%%%%%%%%%%%%%%%%%%%%%%%%%%%%%%%%%%%%%%%%%%%%%%%%%%%%%%%%%%%%%%%%%%%%%%%%%
\section{Unbiased search method \label{theory}}

Our data vector $\hat{P}$ (referred to as the cross-spectral density (CSD) in relevant literature \cite{LVK_isotropic}) is formed from cross-correlations between the strain measured in the different detectors in our network,
    \begin{equation}
        \hat{P}_{IJ}(f,t)=\frac{2}{\tau}\hat{s}_I(f,t)\hat{s}^*_J(f,t),
    \end{equation}
    with $I\ne J$ labelling these detectors, and $\tau$ the length of the time segments used to measure this cross-power (typically $192\,\mathrm{s}$ in anisotropic LVK analyses, with 50\% of the data overlapping~\cite{directional_search}).
Each cross-correlation is formed from data at a particular frequency $f$ (which can be positive or negative) and time segment $t$.
We assume that the strain data, $\hat{s}$, are Gaussian, so that the first and second moments of $\hat{P}$ are
    \begin{equation}
    \label{eq:data-mean}
    \ev*{\hat{P}_{IJ}(f,t)}_N=\sum_{\ell m}\tilde{\gamma}^{\ell m}_{IJ}(f,t)\Omega_{\ell m}(t),\\
    \end{equation}
    \begin{align}
    \begin{split}
    \label{eq:data-covariance}
        &\mathrm{Cov}[\hat{P}_{IJ}(f,t),\hat{P}_{I'J'}(f',t')]_N\\
        &=\delta_{ff'}\delta_{tt'}\qty(\delta_{II'}N_I(f,t)+\sum_{\ell m}\tilde{\gamma}_{II'}^{\ell m}(f,t)\Omega_{\ell m}(t))\\
        &\qquad\qquad\times\qty(\delta_{JJ'}N_J(f,t)+\sum_{\ell'm'}\tilde{\gamma}_{JJ'}^{\ell'm'*}(f,t)\Omega^*_{\ell'm'}(t))\\
        &+\delta_{f,-f'}\delta_{tt'}\qty(\delta_{IJ'}N_I(f,t)+\sum_{\ell m}\tilde{\gamma}_{IJ'}^{\ell m}(f,t)\Omega_{\ell m}(t))\\
        &\qquad\qquad\times\qty(\delta_{I'J}N_J(f,t)+\sum_{\ell'm'}\tilde{\gamma}_{I'J}^{\ell'm'*}(f,t)\Omega^*_{\ell'm'}(t)),
    \end{split}
    \end{align}
    where $\ev{\cdots}_N$ denotes an average over noise realisations, $N_I(f,t)$ is the noise power spectral density PSD in detector $I$, and $\tilde{\gamma}^{\ell m}_{IJ}(f,t)$ are the spherical harmonic components of the overlap reduction function (ORF) of detector baseline $IJ$.
We have absorbed a frequency-dependent factor into the definition of these compared to the definition usually found in the literature, as this simplifies many of the expressions below,
        \begin{equation}
        \label{ORF}
            \tilde{\gamma}^{\ell m}_{IJ}\equiv\frac{3H_0^2}{2\uppi^2f_\mathrm{ref}^3}(f/f_\mathrm{ref})^{\alpha-3}\gamma^{\ell m}_{IJ},
        \end{equation}
    where $\alpha$ is the assumed spectral index of the astrophysical/cosmological signal and $f_{\rm ref}=25\,\mathrm{Hz}$.
This definition factors out the frequency-dependence of the signal, so that the GWB spherical harmonics $\Omega_{\ell m}$ depend only on time, due to different realisations of time-dependent shot noise.
(Here we assume for simplicity that the signal is a power-law in frequency, as expected for an astrophysical background from quasi-circular binary inspirals well before merger; incorporating more detailed modelling of the frequency dependence of the signal may allow for more effective removal of the shot noise~\cite{ValbusaDallArmi:2022htu}.)

We estimate the spherical harmonic components of the GWB as~\cite{Thrane:2009fp}
    \begin{equation}
        \hat{\Omega}_{\ell m}=\sum_{\ell'm'}(\Gamma^{-1})_{\ell m,\ell'm'}\hat{X}_{\ell'm'},
    \label{clean_map}
    \end{equation}
    where the \emph{dirty map} $\hat{X}_{\ell m}$ and \emph{Fisher matrix} $\Gamma_{\ell m,\ell'm'}$ are defined as
    \begin{align}
    \begin{split}
    \label{eq:dirty-map-fisher-matrix-defs}
        \hat{X}_{\ell m}&=(\tilde{\gamma}^{\ell m}|\hat{P}),\\
        \Gamma_{\ell m,\ell'm'}&=(\tilde{\gamma}^{\ell m}|\tilde{\gamma}^{\ell'm'}).
    \end{split}
    \end{align}
Here we have introduced a noise-weighted inner product over time-, frequency-, and baseline-dependent complex functions,
    \begin{equation}
        (A|B)=\sum_{J>I}\sum_{f,t}\frac{A_{IJ}^*(f,t)B_{IJ}(f,t)}{\bar{P}_I(f,t)\bar{P}_J(f,t)}.
    \end{equation}
Note that there is a subtlety associated with the quantity we are using to do the noise weighting, $\bar{P}_I$.
To leading order this is just the noise PSD $N_I$, but since it is measured from the strain data, it will inevitably include some contribution from the GWB signal.
In principle, since this is a data-dependent quantity, we should treat it as a random variable.
However, in practice it is estimated independently of the data segment under consideration (e.g. using neighbouring segments), so that it can be effectively considered as a deterministic quantity.
It is still unclear exactly what value this deterministic noise estimate will take, particularly when we consider corrections of order $\Omega_{\ell m}$.
As we show below, this is a deficiency of the existing search method, which can be rectified in a way that also minimises the impact of shot noise.

To see this, it is necessary first to write down a form of Eqs.~\eqref{eq:data-mean} and~\eqref{eq:data-covariance} which averages over shot noise and cosmological fluctuations, as well as detector noise.
The first and second moments of the GWB spherical harmonics under these averages are~\cite{Jenkins:2019nks}
    \begin{align}
    \label{eq:cosmo-shot-noise-first-moment}
        \ev{\Omega_{\ell m}(t)}_{S,\Omega}&=\delta_{\ell0}\sqrt{4\uppi}\bar{\Omega},\\
    \label{eq:cosmo-shot-noise-second-moment}
        \mathrm{Cov}[\Omega_{\ell m}(t),\Omega_{\ell'm'}(t')]_{S,\Omega}&=\delta_{\ell\ell'}\delta_{mm'}\qty(C_\ell+\delta_{tt'}W_\tau),
    \end{align}
    where $C_\ell$ is the true GWB angular power spectrum that we are aiming to measure, and $W_\tau\gg C_\ell$ is the shot noise power in a segment of length $\tau$, which can be calculated for a given model of the AGWB using expressions in Ref.~\cite{Jenkins:2019nks}.
We see that the shot noise gives an excess contribution to the variance of GWB spherical harmonic measurements made at the same time $t=t'$, and that this contribution is spectrally white (i.e. independent of angular multipole $\ell$).
Combining Eqs.~\eqref{eq:data-mean}, \eqref{eq:data-covariance}, and~\eqref{eq:cosmo-shot-noise-second-moment}, we find
\begin{widetext}
    \begin{align}
    \begin{split}
    \label{eq:data-full-average}
        &\mathrm{Cov}[\hat{P}_{IJ}(f,t),\hat{P}_{I'J'}(f',t')]_{N,S,\Omega}=\sum_{\ell m}\tilde{\gamma}_{IJ}^{\ell m}(f,t)\tilde{\gamma}_{I'J'}^{\ell m*}(f',t')(C_\ell+\delta_{tt'}\mathcal{W}_\tau)\\
        &+\delta_{ff'}\delta_{tt'}\bigg[(\delta_{II'}N_I(f,t)+\tilde{\gamma}_{II'}^{00}(f,t)\sqrt{4\uppi}\bar{\Omega})(\delta_{JJ'}N_J(f,t)+\tilde{\gamma}_{JJ}^{00}(f,t)\sqrt{4\uppi}\bar{\Omega})+\sum_{\ell m}\tilde{\gamma}_{II'}^{\ell m}(f,t)\tilde{\gamma}_{JJ'}^{\ell m*}(f,t)(C_\ell+\mathcal{W}_\tau)\bigg]\\
        &+\delta_{f,-f'}\delta_{tt'}\bigg[(\delta_{IJ'}N_I(f,t)+\tilde{\gamma}_{IJ'}^{00}(f,t)\sqrt{4\uppi}\bar{\Omega})(\delta_{I'J}N_J(f,t)+\tilde{\gamma}_{I'J}^{00}(f,t)\sqrt{4\uppi}\bar{\Omega})+\sum_{\ell m}\tilde{\gamma}_{IJ'}^{\ell m}(f,t)\tilde{\gamma}_{I'J}^{\ell m*}(f,t)(C_\ell+\mathcal{W}_\tau)\bigg].
    \end{split}
    \end{align}
A naive first guess for an appropriate $C_\ell$ estimator is to compute the average power in the $(2\ell+1)$ spherical harmonics corresponding to a given multipole $\ell$,
    \begin{equation}
        \hat{C}_\ell^{\rm (raw)}=\frac{1}{2\ell+1}\sum_m\abs*{\hat{\Omega}_{\ell m}}^2.
    \end{equation}
Using Eqs.~\eqref{eq:dirty-map-fisher-matrix-defs} and~\eqref{eq:data-full-average} we can show that the expected value of this quantity is
    \begin{equation}
        \ev*{\hat{C}_\ell^{\rm (raw)}}_{N,S,\Omega}=C_\ell+\delta_{\ell0}4\uppi\bar{\Omega}^2+\mathcal{N}_\ell.
    \end{equation}
The second term here is to be expected, and simply reflects the fact that the monopole ($\ell=0$) is not zero-mean like the other spherical harmonics.
The third term is a \emph{noise-induced bias}, which is obtained by expanding out the expressions in Eq.~\eqref{eq:dirty-map-fisher-matrix-defs},
    \begin{align}
    \begin{split}
    \label{eq:bias}
        \mathcal{N}_\ell&=\frac{1}{2\ell+1}\sum_m\sum_{\ell'm'}\sum_{\ell''m''}(\Gamma^{-1})_{\ell m,\ell'm'}(\Gamma^{-1})_{\ell''m'',\ell m}\sum_{J>I}\sum_{J'>I'}\sum_{f,t}\sum_{f',t'}\frac{\tilde{\gamma}^{\ell'm'*}_{IJ}(f,t)\tilde{\gamma}^{\ell''m''}_{I'J'}(f',t')}{\bar{P}_I(f,t)\bar{P}_J(f,t)\bar{P}_{I'}(f',t')\bar{P}_{J'}(f',t')}\\
        &\times\delta_{tt'}\Bigg\{\sum_{LM}\tilde{\gamma}^{LM}_{IJ}(f,t)\tilde{\gamma}^{LM*}_{I'J'}(f',t')W_\tau\\
        &\qquad\qquad+\delta_{ff'}\qty[(\delta_{II'}N_I(f,t)+\tilde{\gamma}^{00}_{II'}\sqrt{4\uppi}\bar{\Omega})(\delta_{JJ'}N_J(f,t)+\tilde{\gamma}^{00}_{JJ'}\sqrt{4\uppi}\bar{\Omega})+\sum_{LM}\tilde{\gamma}^{LM}_{II'}\tilde{\gamma}^{LM*}_{JJ'}(C_L+W_\tau)]\\
        &\qquad\qquad+\delta_{f,-f'}\qty[(\delta_{IJ'}N_I(f,t)+\tilde{\gamma}^{00}_{IJ'}\sqrt{4\uppi}\bar{\Omega})(\delta_{I'J}N_J(f,t)+\tilde{\gamma}^{00}_{I'J}\sqrt{4\uppi}\bar{\Omega})+\sum_{LM}\tilde{\gamma}^{LM}_{IJ'}\tilde{\gamma}^{LM*}_{I'J}(C_L+W_\tau)]\Bigg\}.
    \end{split}
    \end{align}
\end{widetext}
This expression involves contributions from pure detector noise, shot noise, and from the signal itself.
In the limiting case where there is no signal and no shot noise, this simplifies to
    \begin{equation}
    \label{eq:bias-leading-order}
        \mathcal{N}_\ell^{(\mathrm{lim})}=\frac{1}{2\ell+1}\sum_m(\Gamma^{-1})_{\ell m,\ell m},
    \end{equation}
    which is the standard expression first derived in Ref.~\cite{Thrane:2009fp}.
(We have assumed that $\bar{P}_I=N_I$ in this case.)

In existing LVK analyses~\cite{directional_search}, this leading-order contribution to the bias is explicitly subtracted from the angular power spectrum estimates, so that the corresponding estimator is given by
    \begin{align}
    \begin{split}
    \label{C_l_curr_biased}
        \hat{C}_\ell^{(\mathrm{curr})}&=\hat{C}_\ell^{\rm (raw)}-\mathcal{N}_\ell^{(\mathrm{lim})}\\
        &=\frac{1}{2\ell+1}\sum_m\qty[\abs*{\hat{\Omega}_{\ell m}}^2-(\Gamma^{-1})_{\ell m,\ell m}].
    \end{split}
    \end{align}
However, it is clearly not possible to set the GWB signal and shot noise to zero in real data, so Eq.~\eqref{eq:bias} tells us that, even if we subtract the usual bias term~\eqref{eq:bias-leading-order}, the current estimator $\hat{C}_\ell^{({\rm curr})}$ is still biased by an amount much larger than the signal we are searching for, since $N_I\bar{\Omega}\gg C_\ell$ and $W_\tau\gg C_\ell$.
The exact size of this bias depends on the properties of the noise estimates $\bar{P}_I(f,t)$, which themselves include further contributions from the GWB.
It is possible in principle that one could define a procedure for generating the noise estimates such that the bias \emph{is} on average given by Eq.~\eqref{eq:bias-leading-order}; however, it is not at all clear how to do this in practice.

A better alternative is to notice that all of the contributions to the bias~\eqref{eq:bias} can be traced back to terms proportional to $\delta_{tt'}$ in Eq.~\eqref{eq:data-full-average}.
In other words, \emph{the bias comes from auto-correlations of each data segment with itself}.
One can thus remove the bias \emph{exactly}, by explicitly excluding such auto-correlations when constructing the $C_\ell$ estimate.
This is the same method that was first suggested in Ref.~\cite{Jenkins:2019nks} for removing the bias induced by shot noise; however, we see here that the same procedure removes all contributions to the bias, including those from detector noise.

Practically, one proceeds by constructing multiple dirty maps $\hat{X}_{\ell m}^{(i)}$, $i=1,\dots,n$, corresponding to disjoint subsets (or epochs) of the total dataset, each with its own Fisher matrix $\Gamma^{(i)}_{\ell m,\ell'm'}$.
The optimal $C_\ell$ estimate is then formed by summing over all nonequal pairs of maps,
    \begin{equation}
        \hat{C}_\ell^{({\rm opt})}=\frac{1}{2\ell+1}\frac{1}{n(n-1)}\sum_m\sum_{i\ne j}\hat{\Omega}^{(i)}_{\ell m}\hat{\Omega}^{(j)*}_{\ell m}.
    \end{equation}
One can then explicitly show, by a very similar calculation to those above, that this is unbiased,
    \begin{equation}
        \ev*{\hat{C}_\ell^{({\rm opt})}}_{N,S,\Omega}=C_\ell+\delta_{\ell0}4\uppi\bar{\Omega}^2.
    \end{equation}
We can calculate the variance of this estimator, using the fact that the spherical harmonic estimates $\hat{\Omega}^{(i)}_{\ell m}$ are Gaussian by the central limit theorem (since they are each formed by summing over many independent data segments).
The general expression is very lengthy, but in the \textit{weak-signal regime} where $\Gamma^{-1}\gg C_\ell,\,\bar{\Omega}^2,\,W_\tau$ it simplifies to give
    \begin{align}
    \begin{split}
    \label{eq:C_ell-variance}
        \mathrm{Var}[\hat{C}_\ell^{({\rm opt})}]&\simeq\frac{2}{(2\ell+1)^2}\frac{1}{n^2(n-1)^2}\\
        &\times\sum_{m,m'}\sum_{i\ne j}(\Gamma^{-1})^{(i)}_{\ell m,\ell m'}(\Gamma^{-1})^{(j)}_{\ell m',\ell m}.
    \end{split}
    \end{align}
Let us consider the simplest case, where the Fisher matrices for each of the subsets of the data are equal to one another, and are thus given by $1/n$ times the total Fisher matrix,
    \begin{equation}
        \Gamma^{(i)}_{\ell m,\ell'm'}\simeq\frac{1}{n}\Gamma_{\ell m,\ell'm'}.
    \end{equation}
We can then simplify Eq.~\eqref{eq:C_ell-variance} to give
    \begin{equation}
        \mathrm{Var}[\hat{C}_\ell^{({\rm opt})}]\simeq\frac{2}{(2\ell+1)^2}\frac{n}{n-1}\sum_{m,m'}\abs*{(\Gamma^{-1})_{\ell m,\ell m'}}^2.
    \label{theor_unc_opt}
    \end{equation}
This is equal to the variance found in Ref.~\cite{Thrane:2009fp} for the current estimator, times a factor of $n/(n-1)$.
We therefore see that our unbiased estimator is equally efficient in the limit of large $n$, just as in Ref.~\cite{Jenkins:2019nks}.

%%%%%%%%%%%%%%%%%%%%%%%%%%%%%%%%%%%%%%%%%%%%%%%%%%%%%%%%%%%%%%%%%%%%%%%%%%%%%%
\section{Validation on mock data \label{mock}}
\subsection{O3 sensitivity \label{O3}}
\subsubsection{Pipeline \& settings}

To run our anisotropic analysis, we make use of the publicly available code \texttt{PyStoch}~\cite{Ain:2018zvo, Suresh:2020khz}. This pipeline has been designed and tested on O3 folded data~\cite{folding} (CSD \& PSD), for which it computes the dirty map and Fisher matrix either in pixel space or spherical harmonic space. The user specifies an assumed spectral index, $\alpha$; a frequency range (the whole GWB frequency range in the standard LVK analysis corresponds to $[20,1726]$ Hz); a detector baseline; and the number of pixels or maximum multipole they want to probe, $\ell_\mathrm{max}$.

In our case, we adopt $\alpha={2}/{3}$ to consider an astrophysical background~\cite{phinney2001practical} and run \texttt{PyStoch} in spherical harmonic space assuming the LIGO Hanford-Livingston (HL) baseline. We produce mock CSDs\footnote{In this exercise, only the CSDs are drawn randomly to represent individual data realisations. The PSD for each realisation is kept fixed, essentially reducing the Fisher matrix to a known weight in order to model the ideal scenario of equal-length datasets. This approach simplified computations.} made up of the following components:
\begin{itemize}
    \item \textit{Detector noise}. This is generated as random Gaussian realisations (different at every element in our time-frequency CSD array) of the folded data~\cite{folded_data} at O3 sensitivity. For increased LIGO sensitivity (see section \ref{A+}), we use the expected amplitude spectral densities (ASD's) from Ref.~\cite{2018LRR....21....3A}.
    \item \textit{Shot noise}. We generate shot-noise sky maps (different at every time segment in our data) with \texttt{Healpy} \cite{Healpy, HEALPix} as random realisations of the same flat $C_\ell$ spectrum (corresponding to white noise). We then multiply each map with the ORF (given by Eq.~\ref{ORF}) to produce a shot-noise CSD.
    \item \textit{Mock AGWB signal}. We simulate a sky map that represents the AGWB signal. Again, this is done with \texttt{Healpy}, using a model of the cosmological $C_\ell$ spectrum from Ref.~\cite{PhysRevD.100.063508} (see section \ref{shot+LSS}). We subsequently multiply this map with the ORF (Eq.~\ref{ORF}) to produce a CSD for the GWB.
\end{itemize}

We produce dirty maps, Fisher matrices and clean maps from our simulated CSDs by means of Eqs.~(\ref{clean_map}) and (\ref{eq:dirty-map-fisher-matrix-defs}). In the results that follow, we have not applied any regularisation schemes to the Fisher matrix inversion. See \hyperref[appendix_b]{Appendix B} for a discussion on the impact of regularisation in our study. %
Given that we are working in the weak-signal regime, we expect the associated statistical uncertainty in the $C_\ell$ calculation to be large compared to the injected values for the short-duration ($\sim1\,\mathrm{yr})$ datasets simulated here.

Our \texttt{PyStoch}-based analysis is sufficiently fast that we can simulate many independent realisations of shot noise and detector noise.
We achieve this by choosing appropriate values for the maximum spherical multipole ($\ell_\mathrm{max}=8$) and the analysed frequency range ($[20,520]\,\mathrm{Hz}$), which allow us to perform our simulations relatively cheaply while still capturing the key features of the problem.
We stress however that our methods are general, and can be applied to a broader range of multipoles and frequencies in a full analysis.

\begin{figure}
\centering
\includegraphics[width=\linewidth]{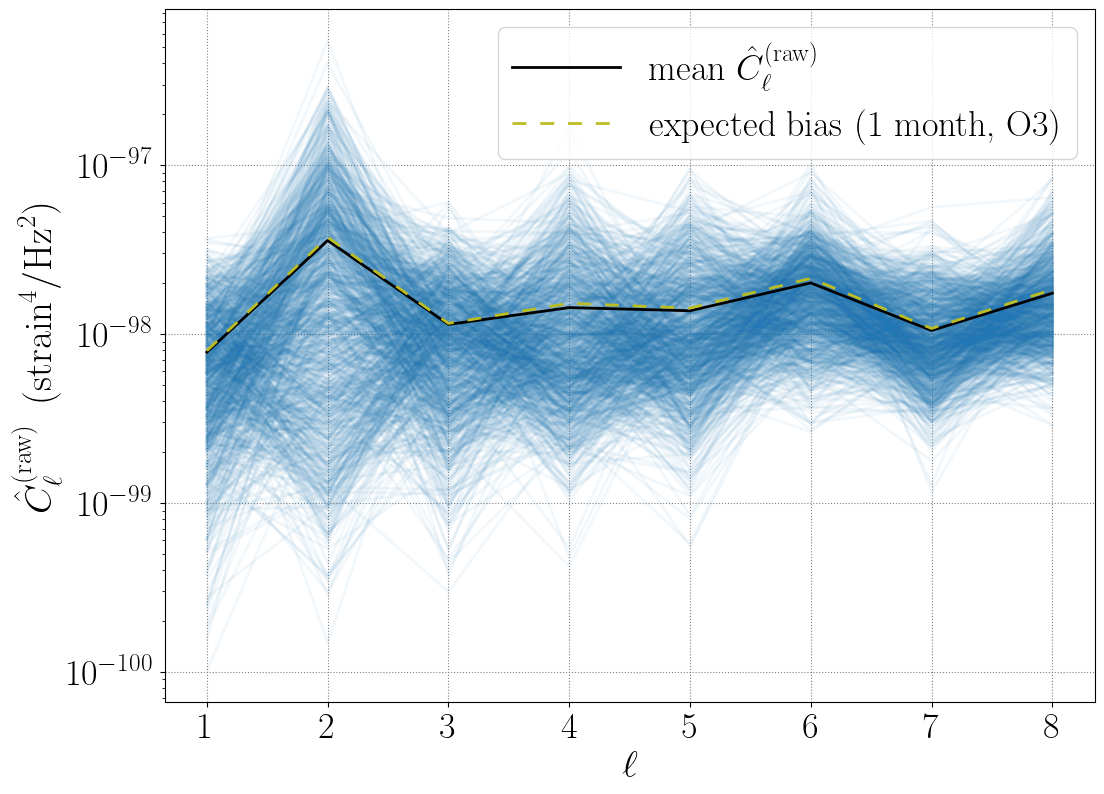}
\caption{The angular power spectra of 1,000 clean maps (blue curves), as recovered from simulations with detector noise only at O3 sensitivity (mock 1-month data sets). The mean $\hat{C}_\ell^{\rm (raw)}$ spectrum (black curve) matches the expected bias in the weak-signal limit, $\mathcal{N}_\ell^{(\mathrm{lim})}$ (dashed yellow curve).}
\label{all_biased_det_noise}
\end{figure}

%\subsubsection{Detector-noise bias subtraction}
\subsubsection{Validation with existing estimator}
\begin{figure}
\centering
\includegraphics[width=\linewidth]{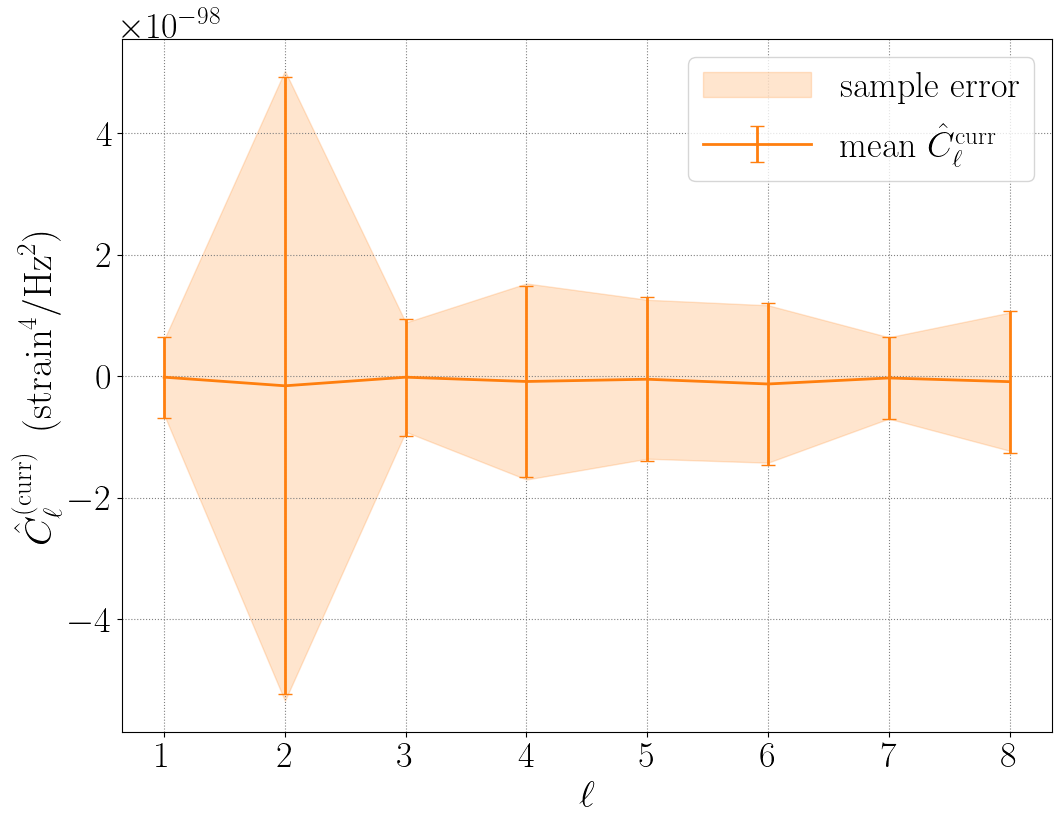}
\caption{The mean $\hat{C}_\ell^{({\rm curr})}$ estimator of 1,000 simulations with detector noise only, at O3 sensitivity. We plot the predicted variance of the estimator with error bars, and the sample variance in our ensemble of simulations as a shaded region.}
\label{mean_unbiased_stand_det_noise}
\end{figure}

First we confirm that the existing estimator ($\hat{C}_\ell^{({\rm curr})}$) performs as expected in the weak-signal regime by running \texttt{PyStoch} on simulated detector noise, without injecting shot noise or an underlying cosmological signal. In Fig.~\ref{all_biased_det_noise}, we plot the clean-map angular power spectra of $1,000$ random detector noise realisations (blue curves), along with their mean (black curve).
As expected, this mean curve exhibits excellent agreement with the weak-signal limit of the bias, $\mathcal{N}_\ell^{(\mathrm{lim})}$, as given by Eq.~(\ref{eq:bias-leading-order}) (dashed yellow curve).
In Fig. \ref{mean_unbiased_stand_det_noise} we plot the estimator $\hat{C}_\ell^{(\mathrm{curr})}$, which corresponds to the same spectra but with the weak-signal limit of the bias subtracted. At O3 sensitivity, the shot-noise and GWB contributions to the bias in the current estimator are much smaller than the uncertainty of the estimate, so that the resulting curve is consistent with zero for pure detector noise. In the same figure, we plot the expected variance of this current estimator from our calculations (Eq.~(\ref{theor_unc_opt}) in the $n\rightarrow\infty$ limit), as well as the sample variance observed in our ensemble of $1,000$ noise realisations.
The excellent agreement between these is a useful consistency check for our simulation pipeline.

%\subsubsection{Shot noise \& AGWB estimation \label{shot+LSS}}
\subsubsection{Modelling shot noise and large-scale structure \label{shot+LSS}}
We now turn to the full problem, in which the simulated data contain a GWB signal with intrinsic angular correlations due to cosmic large-scale structure, as well as spurious correlations due to temporal shot noise.
We simulate the maps that we inject in random detector noise realisations using realistic values for the shot noise and LSS (monopole and anisotropies) cited in Ref.~\cite{PhysRevD.100.063508}. The reported values in that work for the angular power spectrum are in units of $\Omega_{\rm GW}^2$ at a frequency of $f_\mathrm{ref}=65$ Hz. We consider the shot noise and LSS values estimated for a subtracted foreground of $r_*=250$ Mpc (some degree of foreground subtraction must be assumed in order to avoid a divergence in the expected shot noise power at short distances). See \hyperref[appendix_a]{Appendix A} for a detailed conversion of these results into the right units and $f_{\rm ref}$ for our analysis.

\begin{figure}
\centering
\includegraphics[width=\linewidth]{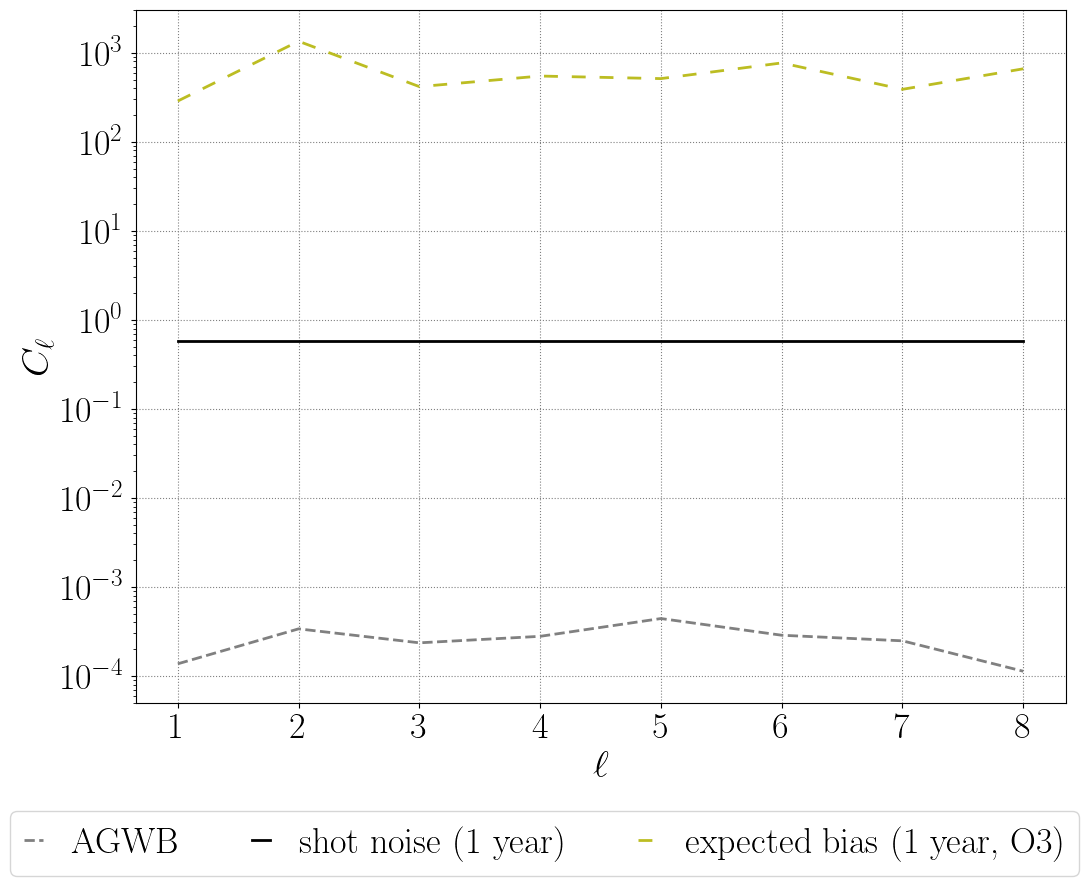}
\caption{Expected angular power spectrum of the temporal shot noise (black curve) and the AGWB (dashed grey curve). The dashed yellow curve denotes the expected bias in the weak-signal limit, given by the O3 (HL) detector noise. Both the detector noise and the shot noise correspond to 1 year of data. All curves have been normalised with respect to the AGWB monopole.}
\label{exp_noise+signal}
\end{figure}

In Fig. \ref{exp_noise+signal}, we plot the predicted bias for the existing estimator $\hat{C}_\ell^{(\mathrm{curr})}$ in the weak-signal limit, as given by Eq. (\ref{eq:bias-leading-order}), along with the expected angular power spectra for the shot noise and underlying AGWB. Everything has been normalised with respect to the monopole, $C_\ell\to C_\ell/(4\pi\bar{\Omega}^2)$. We note that for the detector noise and the shot noise, the assumed amount of time over which each data set has been collected is one year. For O3 sensitivity, the detector noise is orders of magnitude above the shot noise and AGWB signal, so that we are unambiguously in the weak-signal regime. It is also clear that for O3 sensitivity, we cannot expect to probe the angular power spectrum of the AGWB. This is consistent with the non-detection of such a signal in O3 data~\cite{directional_search}.

\begin{figure}
\centering
\includegraphics[width=\linewidth]{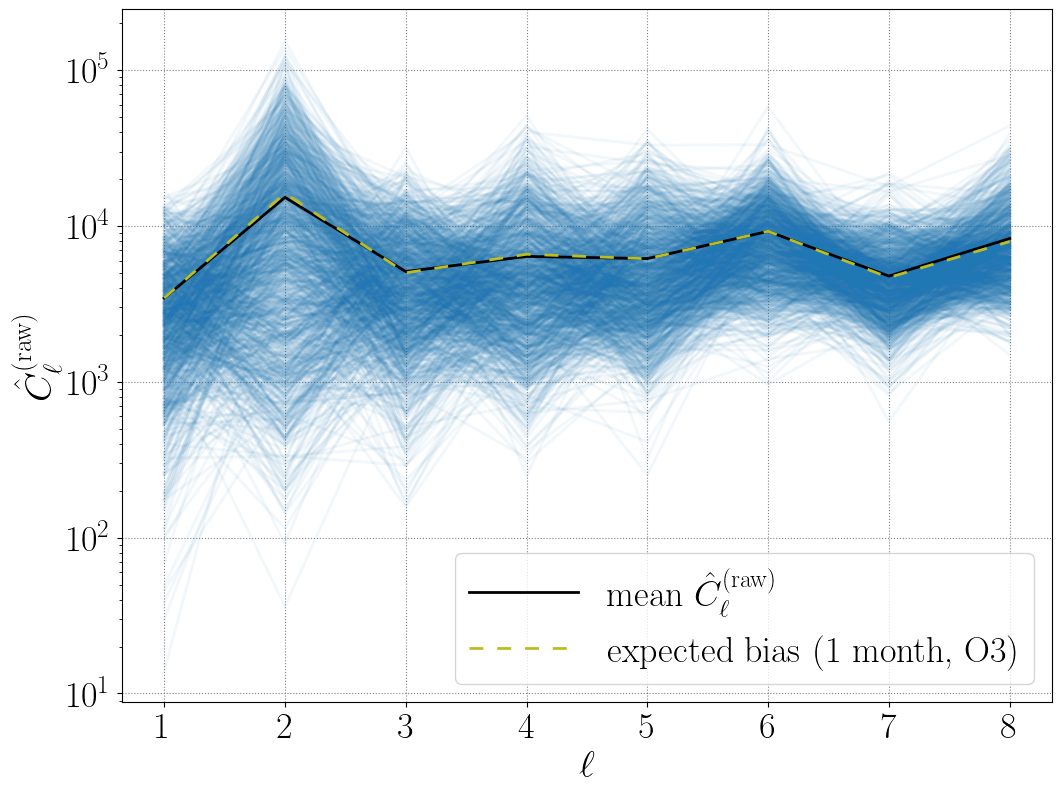}
\caption{The angular power spectra of 1,200 clean maps (blue curves), normalised to the monopole, as recovered from simulations with O3 detector noise, shot noise and the AGWB (mock 1-month data sets). The mean $\hat{C}_\ell^{\rm (raw)}$ spectrum (black curve) matches the expected (O3) bias in the weak-signal regime (dashed yellow curve).}
\label{all_biased}
\end{figure}

\begin{figure}
\centering
\includegraphics[width=\linewidth]{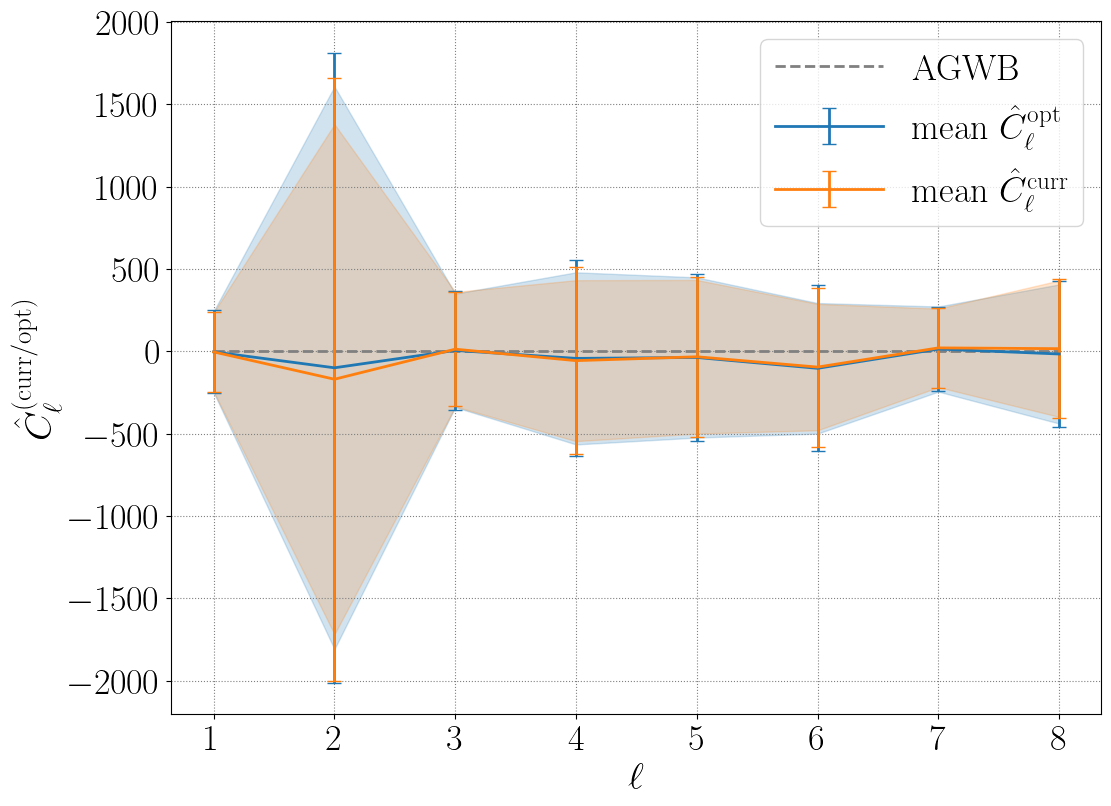}
\caption{Comparison of the performance of the $\hat{C}_\ell^{({\rm curr})}$ (orange) and $\hat{C}_\ell^{({\rm opt})}$ (blue) estimators in a realistic O3-like scenario. We plot the expected variances as error bars and the sample variances as error bands. Everything has been normalised with respect to the AGWB monopole. For 1 year of O3 data, the AGWB (dashed grey curve) is very small compared to the uncertainties in the angular power spectrum estimate.}
\label{estim_comp}
\end{figure}

\subsubsection{Injected map simulations and angular power spectra}
We generate detector noise and shot noise assuming one-month data sets, such that both vary randomly at every time segment in our data sets. We then inject the same, fixed underlying LSS signal to each data set.

In total, we produce 1,200 clean maps corresponding to one month of data each. In Fig. \ref{all_biased}, we plot the angular power spectra for these clean maps, along with the ensemble mean. This once again matches the bias in Eq.~\eqref{eq:bias-leading-order}, as expected in the weak-signal regime.

We subsequently distribute the clean maps in 100 sets of 12 maps each to construct one-year data sets\footnote{The corresponding dirty map and Fisher matrix for each one-year dataset that are used in the $\hat{C}_\ell^{({\rm curr})}$ estimator calculation are obtained by adding up the 12 different dirty maps and Fisher matrices, respectively.}. For each set, we calculate the $\hat{C}_\ell^{({\rm curr})}$ and $\hat{C}_\ell^{({\rm opt})}$ estimators. The results are plotted in Fig. \ref{estim_comp}, along with the predicted variance (given by Eq. (\ref{theor_unc_opt}) for $\hat{C}_\ell^{({\rm opt})}$, and the same expression in the $n\rightarrow\infty$ limit for $\hat{C}_\ell^{({\rm curr})}$) and sample variance for both estimators, which are in good agreement in both cases.
Due to the weak signal and short timespan (relative to O3 sensitivity), both estimators have large uncertainties, and are both equally consistent with the injected signal.
(The variance of the unbiased estimator is larger by a factor of 12/11, as expected; this factor can be reduced by dividing the data into a larger number of subsets.)
This confirms that the effects of shot noise and the bias in the existing analysis method are both negligibly small at O3 sensitivity.
In the section below, we show that this will no longer be true at design sensitivity.

\begin{figure}
\centering
\includegraphics[width=\linewidth]{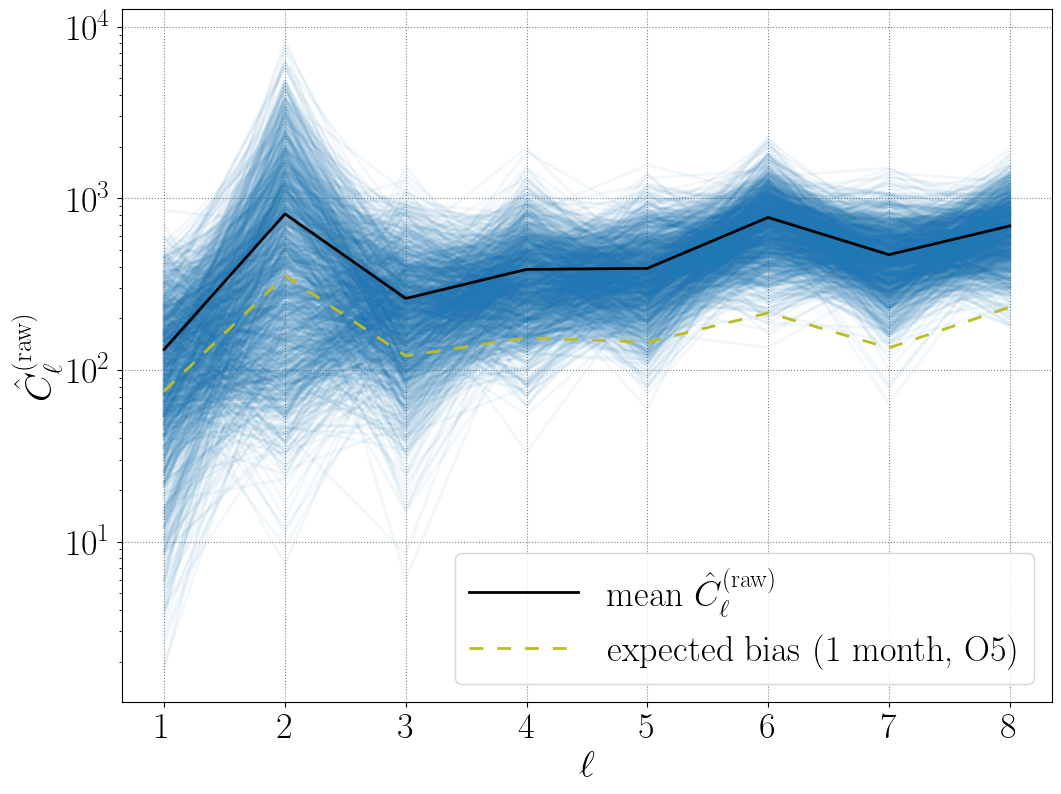}
\caption{The angular power spectrum of 1,200 clean maps (blue curves), normalised to the monopole, as recovered from simulations with O5 detector noise, shot noise and the AGWB (mock 1-month data sets). The mean $\hat{C}_\ell^{\rm (raw)}
$ spectrum (black curve) and the expected (O5) bias in the weak-signal regime (dashed yellow curve) are also plotted. The discrepancy between the black and yellow curves indicates the presence of bias due to beyond-weak-signal-limit effects.}
\label{all_biased_O5}
\end{figure}

\begin{figure}
\centering
\includegraphics[width=\linewidth]{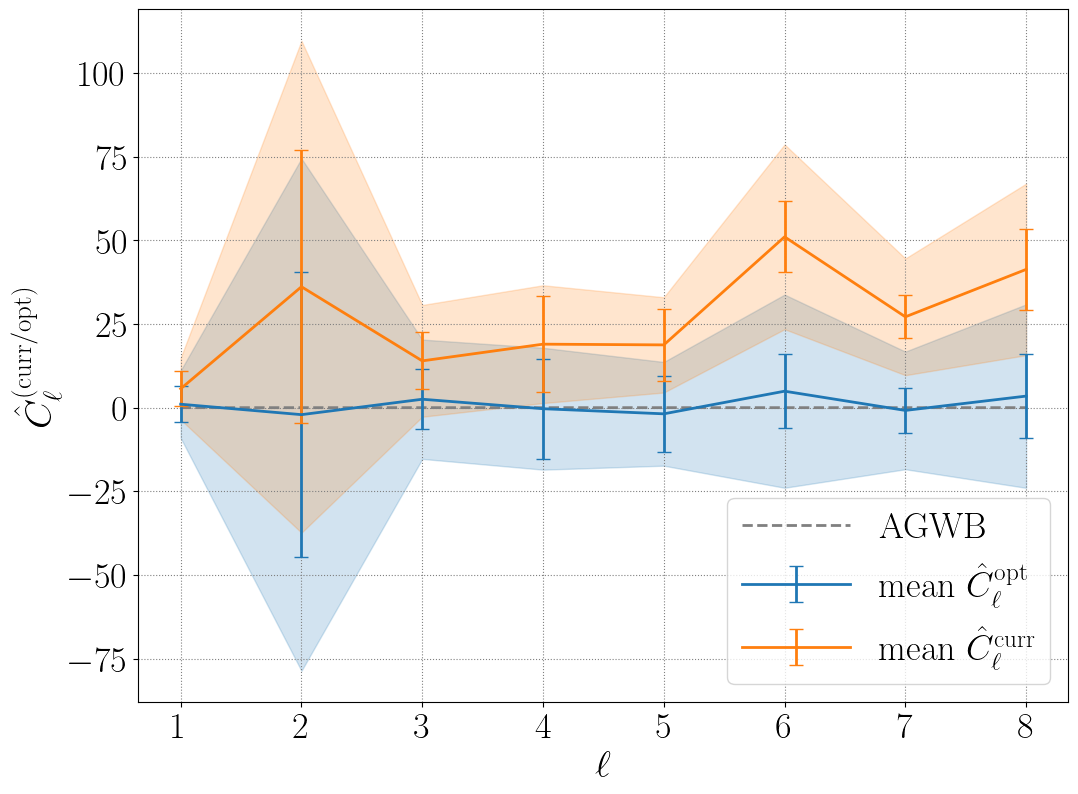}
\caption{Comparison of the performance of the $\hat{C}_\ell^{({\rm curr})}$ (orange curve) and $\hat{C}_\ell^{({\rm opt})}$ (blue) estimators. We plot the expected variances as error bars and the sample variances as error bands. Everything has been normalised with respect to the AGWB monopole. For 1 year of O5 data, the AGWB spectrum (dashed grey curve) is still small compared to the uncertainties in the angular power spectrum calculation.}
\label{estim_comp_O5}
\end{figure}

\subsection{O5 sensitivity \label{A+}}
Advanced LIGO is expected to reach design sensitivity, A+, during the O5 observing run~\cite{design_document}. We run simulations for this detector sensitivity to see how the two $C_\ell$ estimators perform in this setting. We follow the procedure that is outlined in section \ref{O3}, with the sole difference that the detector noise is generated for A+ instead of O3 sensitivity.

Again, we produce 1,200 clean maps corresponding to one month of data each. We plot the angular power spectra of all clean maps in Fig. \ref{all_biased_O5}. This time, their mean clearly lies above the weak-signal expression for the bias, $\mathcal{N}_\ell^{(\mathrm{lim})}$. This is unsurprising, as the monopole and shot noise terms in Eq.~\eqref{eq:bias} are significantly larger compared to the detector noise at A+ sensitivity than at O3 sensitivity. As a result, the limiting-case expression for the expected bias is no longer sufficient to describe the actual bias in the analysis. Therefore, the estimator $\hat{C}_\ell^{(\mathrm{curr})}$ we obtain after subtracting the detector bias will still be biased by the sub-leading contributions, which are no longer negligible. In contrast, the optimal estimator introduced in this work is always unbiased; hence, we expect the corresponding curves to lie below those of the current estimator for these simulations.

As before, we distribute the clean maps in 100 sets of 12 maps each and plot the current and optimal estimators in Fig.~\ref{estim_comp_O5}, along with their expected and sample variances.
This time, there is a lack of agreement between the sample variance for each estimator and the theoretical prediction given by Eq.~\eqref{theor_unc_opt}, with the latter being consistently smaller. This can be understood as being due to the breakdown of the weak-signal approximation; one now needs to account for additional contributions to the variance coming from the shot noise and the AGWB.
We also see that the optimal estimator $\hat{C}_\ell^{(\mathrm{opt})}$ returns a spectrum that is, on average, significantly below that of the current estimator $\hat{C}_\ell^{(\mathrm{curr})}$, and is thus closer to the injected $C_\ell$ signal.
This reflects the bias inherent to the current estimator beyond the weak-signal regime, and illustrates that it will be crucial to use the unbiased estimator $\hat{C}_\ell^{(\mathrm{opt})}$ once A+ sensitivity is reached.

\begin{figure*}
    \centering
    \includegraphics[width=\textwidth]{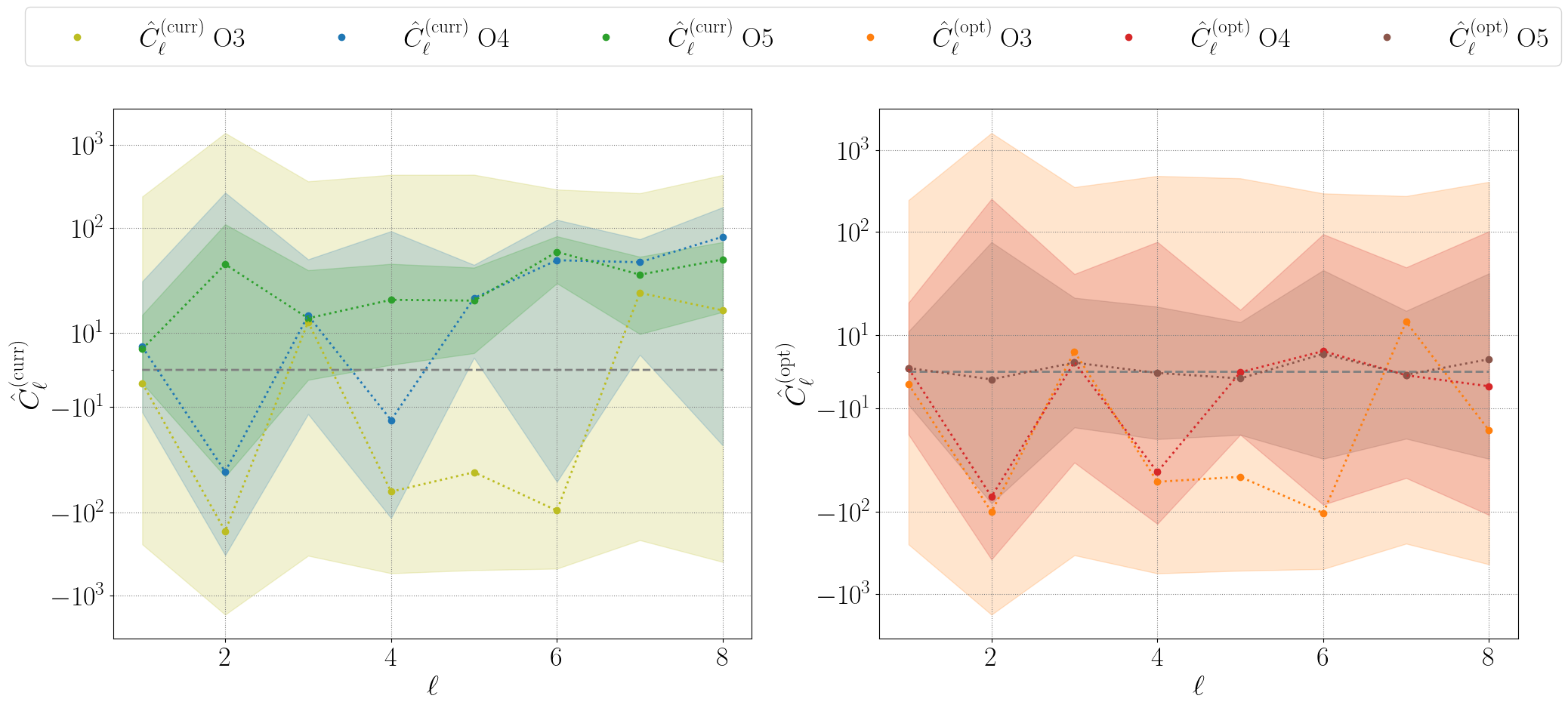}
    \caption{Summary of the $\hat{C}_\ell^{({\rm curr})}$ (left panel) and $\hat{C}_\ell^{({\rm opt})}$ (right panel) estimator performance for the O3, O4 and O5 sensitivities. The sample variances are plotted as error bands. The dashed grey curve denotes the AGWB. It is clear that increasing the detector sensitivity results in the optimal estimator recovering the true AGWB spectrum more precisely, and the bias in the current estimator becoming more evident. Note: because of the symlog scale for the y-axis, error bands with the same physical size but at different vertical positions in the plot span different ranges of values.}
    \label{fig:comparison}
\end{figure*}

\begin{table*}
\centering
\begin{tabular}{ |c|c|c|c|c|c|c|c|c| }
 \hline
 \multicolumn{9}{|c|}{\textbf{Sample STD difference}} \\
 \hline
 & \multicolumn{8}{|c|}{\textbf{$\ell$ mode}} \\
 \hline
\textbf{Detector sensitivity} & 1 & 2 & 3 & 4 & 5 & 6 & 7 & 8\\
 \hline
O3 & -0.015 & -0.109 &  0.037 & -0.116 & -0.071 & -0.25  &  0.088 & 0.039\\
\hline
O4 & 0.352 & -0.11  &  0.552 & -0.131 &  1.204 &  0.484 &  1.12  & 0.791\\
\hline
O5 & 0.612 & 0.492 & 0.838 & 1.078 & 1.318 & 1.85  & 1.556 & 1.612\\
\hline
\end{tabular}
\caption{The difference between the mean of the $\hat{C}_\ell^{({\rm curr})}$ estimator and the AGWB signal at each $\ell$ mode in units of the sample standard deviation (STD).}
\label{sigma_dev}
\end{table*}

\subsection{Discussion}
We close this section with a comparison of our mock simulations for O3, O4, and O5 sensitivity. We repeat our analysis for the projected O4 sensitivity as well to include a forecast of the effect on present-day data. Much like for the case of A+, we notice that the mean $C_\ell$ curve for the 1,200 produced clean maps lies above the detector bias. However, the discrepancy is not so clear for O4, starting as negligible at lower multipoles, and only becoming measurable at higher multipoles. Looking at the performance of the two estimators, we therefore find that the optimal estimator performs visibly better than the current one only for $\ell\gtrsim5$.

In Fig.~\ref{fig:comparison}, we summarise our results for the current (left panel) and optimal (right panel) estimators, for the three different detector sensitivities that we assumed during our analysis. The different $C_\ell$ estimators are plotted as dotted curves of different colours (summarised in the figure legend). Again, the error bands denote the sample variances, while the AGWB (dashed grey curve) is also plotted.

For both estimators, increasing the detector sensitivity results in narrower error bands, as expected. In the case of the optimal estimator, these bands are always centered on the injected AGWB signal; however, for the current estimator, they deviate away from the AGWB, towards positive values (particularly for higher multipoles). Hence, while the optimal estimator yields more precise and accurate $C_\ell$ measurements for increased detector sensitivity, the bias in the current estimator becomes increasingly evident.

We quantify the deviation of the mean of the $\hat{C}_\ell^{({\rm curr})}$ estimator from the AGWB signal in terms of the sample standard deviations difference between them, which is summarised in Table~\ref{sigma_dev}.

%%%%%%%%%%%%%%%%%%%%%%%%%%%%%%%%%%%%%%%%%%%%%%%%%%%%%%%%%%%%%%%%%%%%%%%%%%%%%%
\section{Summary and outlook \label{summary}}
In this study, we have investigated methods for measuring GW background anisotropies in the presense of both detector noise and temporal shot noise. We started by describing the $C_\ell$ estimator currently used in LIGO-Virgo-KAGRA analyses, and demonstrated that this estimator is biased due to contributions from the shot noise and from the target GW signal itself. We then presented a new estimator which avoids this bias by only cross-correlating between data segments at unequal times, validating our predictions on simulations using mock data. The two estimators produce similar results for O3 sensitivity, where we confidently lie in the weak-signal regime. However, we find that for LIGO A+ sensitivity it is crucial to use our new estimator to obtain accurate inferences of the GWB angular power spectrum, with some improvement over the current estimator already visible at O4 sensitivity.

In the future, we plan to extend this work by testing both estimators on real O4 data. Without the computational cost of running the analysis on thousands of simulated data sets, it will be possible to probe higher spherical harmonic $\ell$ modes, in the whole GWB frequency range ($[20,1726]$ Hz). It will also be straightforward to include more detectors in the analysis---such as Virgo and KAGRA---and perform the analysis on multiple detector baselines. This will improve the conditioning of the Fisher matrices involved in the angular power spectrum estimation, allowing access to higher angular resolution.

In addition, it will be interesting to test the capabilities of planned `third-generation' detectors in probing GW anisotropies. It is already clear that the large bias in the angular power spectrum for future detectors will render the current search method unsuitable. On the other hand, the optimal estimator will be able to measure the LSS anisotropies with precision orders of magnitude better than for current detectors.

\section*{Acknowledgments}
We would like to thank Shivaraj Kandhasamy for carefully reviewing this manuscript (LIGO-Document number P2300434) as a part of the LVK collaboration's internal review process. This material is based upon work supported by NSF's LIGO Laboratory, which is a major facility fully funded by the National Science Foundation. The authors acknowledge computational resources provided by the LIGO Laboratory and supported by NSF Grants PHY-0757058 and PHY-0823459. We thank Deepali Agarwal and Stavros Venikoudis for useful comments, and in particular Jishnu Suresh for many useful discussions on how to run \texttt{PyStoch}. NK~is supported by King's College London through an NMES Funded Studentship. %
ACJ is supported by the Science and Technology Facilities Council (STFC) through the UKRI Quantum Technologies for Fundamental Physics Programme [grant number ST/T005904/1]. %
AIR is supported by the European Union's Horizon 2020 research and innovation programme under the Marie Skłodowska-Curie grant agreement No 101064542, and acknowledges support from the NSF award PHY-1912594. %
JDR acknowledges support from National Science Foundation (NSF) Grant No. PHY-2207270. %
MS acknowledges support from the Science and Technology Facility Council
(STFC), UK, under the research grant ST/X000753/1. %
This work was partly enabled by the UCL Cosmoparticle Initiative.

\section*{Appendix A: Unit Conversion \label{appendix_a}}
The GW energy density is related to the GW strain power by~\cite{PhysRevD.59.102001}
\begin{equation}
    \Omega_{\rm GW}(f,\theta) = \frac{2\pi^2}{3 H_0^2} f^3 P(f,\theta),
\end{equation}
where $H_0=67.4$ km ${\rm s^{-1}}$ ${\rm Mpc^{-1}}$~\cite{2020A&A...641A...6P}.
The GW energy density and power can then be factored into spectral and angular terms,
% \begin{equation}
% \begin{gathered}
%     \Omega_{\rm GW}(f,\theta) =\Omega_{\rm GW}(\theta) \bigg(\frac{f}{f_{\rm ref}}\bigg)^\alpha, \\
%     P(f,\theta) = H(f) P(\theta),
% \end{gathered}
% \label{omega}
% \end{equation}
\begin{align}
\Omega_{\rm GW}(f,\theta) &=\Omega_{\rm GW}(\theta) \bigg(\frac{f}{f_{\rm ref}}\bigg)^\alpha,\label{omega}\\
P(f,\theta) &= P(\theta) H(f),
\end{align}
where $H(f) = (f / f_{\rm ref})^{\alpha-3}$. This leads to
\begin{equation}
    P(\theta)=A{\hspace{1mm}}\Omega_{\rm GW}(\theta),
\end{equation}
where $A=3 H_0^2 / (2 \pi^2 f_{\rm ref}^3)$ (this should be further reduced by a factor of 2 to account for taking the one-sided spectrum).

This expression allows us to convert a sky map from units of $\Omega_{\rm GW}$ to units of $P$. Finally, we convert the results in Ref.~\cite{PhysRevD.100.063508} (which correspond to a frequency of 65 Hz) to the appropriate reference frequency by means of Eq.~(\ref{omega}).

\begin{figure*}
\centering
\includegraphics[width=\linewidth]{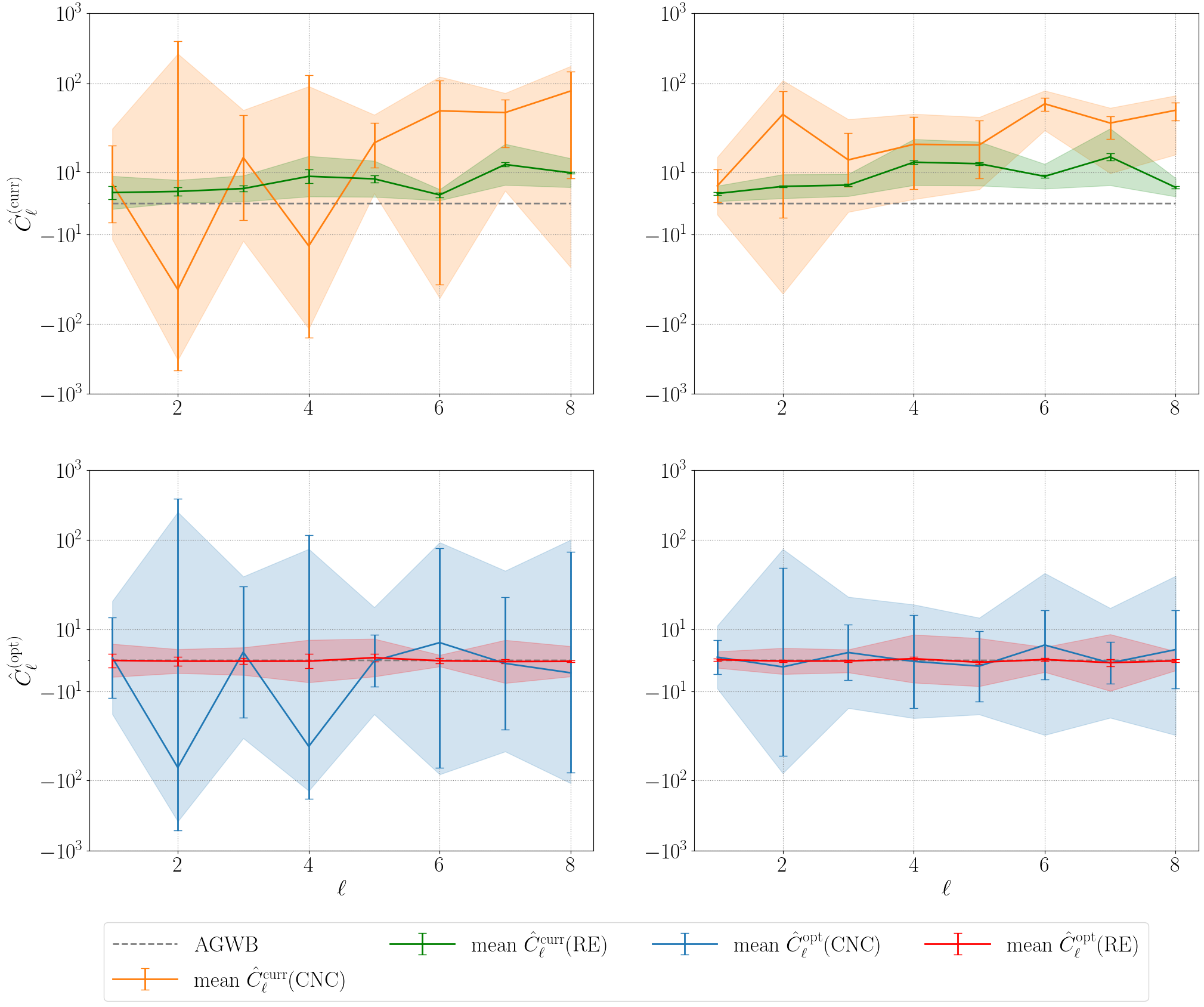}
\caption{Performance of the $\hat{C}_\ell^{({\rm curr})}$ (top) and $\hat{C}_\ell^{({\rm opt})}$ (bottom) estimators, for O4 (left) and O5 (right) detector sensitivity and the CNC and RE regularisation schemes. We plot the expected variances as error bars and the sample variances as error bands. Everything has been normalised with respect to the AGWB monopole. Imposing a more aggressive regularisation scheme results in the bias in the $\hat{C}_\ell^{({\rm curr})}$ estimator becoming more evident.}
\label{regularised_results}
\end{figure*}

\section*{Appendix B: Impact of Regularisation \label{appendix_b}}
In this paper, we have chosen not to employ a regularisation scheme when inverting the Fisher matrix, as the choice for regularisation is not obvious and the community has yet to reach a consensus on the appropriate way to regularise the matrix for present-day detector networks (see e.g. studies presented in~\cite{PhysRevD.107.122002, Erik_reg, PhysRevD.100.043541, PhysRevD.104.123018}). As shown in~\cite{Erik_reg}, the choice of regularisation scheme depends on whether the search prioritises signal detection versus characterisation; this goes beyond the scope of our work, and we do not expect this to qualitatively change the results of this case study. %
Nevertheless, for completeness, we assess the impact that regularising the Fisher matrix has on our results in this appendix. %
We consider two different regularisation schemes on the Fisher matrix inversion:
\begin{itemize}
    \item a Condition Number Cut-off (CNC) of $10^{-5}$ (as done in~\cite{Renzini2018, Renzini2019A, Renzini2019B, Contaldi:2020rht}),
    \item a Reassignment of $1/3$ of the Fisher matrix Eigenvalues (RE) to infinity (as customarily done in LVK collaboration efforts \cite{
PhysRevLett.118.121102, PhysRevD.100.062001, directional_search}).
\end{itemize}
This CNC approach is generally less aggressive than the RE, as it allows for no eigenvalues to be discarded during matrix inversion, while the RE approach blindly discards $1/3$ of the eigenvalues, no matter their actual amplitude. %
The latter can lead to an earlier detection but can also underestimate the measurement errors, while the former can lead to a large amount of noise leaking into the measurement. %
When employing a regularisation scheme, formally the inverse of $\Gamma$, $\Gamma^{-1}$, should be replaced by the regularised inverse $(\Gamma')^{-1}$ . Specifically, this should be done in Eqs.~(\ref{clean_map}, \ref{eq:bias}--\ref{C_l_curr_biased}, \ref{eq:C_ell-variance}, \ref{theor_unc_opt}).

We find that the ability of the current estimator to recover the AGWB signal (first shown in Fig. \ref{mean_unbiased_stand_det_noise}) is not affected by our regularisation schemes. %
Similarly, the performance of the current estimator in the presence of detector noise, shot noise, and the AGWB is not affected significantly in the case of O3 sensitivity (first tested in Fig. \ref{estim_comp}). %
However, the result changes significantly for O4 and O5 sensitivity. %
In Fig.~\ref{regularised_results}, we plot the current (top) and optimal (bottom) estimator, for O4 (left) and O5 (right) sensitivity and the CNC and RE regularisation schemes. %
The CNC scheme has an almost negligible impact on the Fisher matrix inversion and the current and optimal estimators (same for O3), giving almost the same results as reported in Section \ref{mock}. %
This is because the chosen CNC is very close to the Fisher matrix's condition number in this case. %
On the other hand, imposing the RE scheme significantly affects the Fisher matrix inversion and results in an enhancement of the bias in the current estimator. %
As for the optimal estimator, this performs equally well for both regularisation schemes. %
Furthermore, the discrepancy among the theoretical uncertainties and sample variances for O4 and O5 sensitivity appears greater for the RE scheme. %
We conclude that when using a more aggressive regularisation scheme, error estimates are reduced and hence the noise-bias becomes more evident in the current estimator, and the weak-signal approximation looses validity.

%%%%%%%%%%%%%%%%%%%%%%%%%%%%%%%%%%%%%%%%%%%%%%%%%%%%%%%%%%%%%%%%%%%%%%%%%%%%%%
\bibliography{shot-noise.bib}
\end{document}